\documentstyle[epsfig,twocolumn,prb,aps]{revtex}

\begin{document}
\draft

\twocolumn[\hsize\textwidth\columnwidth\hsize\csname @twocolumnfalse\endcsname

\title{Low-energy properties and magnetization plateaus in a 2-leg mixed spin ladder}
\author{ Jizhong Lou and  Changfeng Chen}
\address{
Department of Physics, University of Nevada, Las Vegas, Nevada 89154}
\author{Shaojin Qin}
\address{Department of Physics, Kyushu University, Hakozaki, Higashi-ku,
        Fukuoka 812-8581, Japan \\
{\it and} Institute of Theoretical Physics, P. O. Box 2735, Beijing 100080, P. R. China}
\date{ \today }
\maketitle

\begin{abstract}
Using the density matrix renormalization group technique we investigate the
low-energy properties and the magnetization plateau behavior in a 2-leg mixed spin 
ladder consisting of a spin-1/2 chain coupled with a spin-1 chain.
The calculated results show that the system is in the same universality class as the 
spin-3/2 chain when the interchain coupling is strongly ferromagnetic, but the
similarity between the two systems is less clear under other coupling conditions.
We have identified two types of magnetization plateau phases.  The calculation of
the magnetization distribution on the spin-1/2 and the spin-1 chains on the ladder
shows that one plateau phase is related to the partially magnetized valence-bond-solid 
state, and the other plateau state contains strongly coupled S=1 and s=1/2 spins on
the rung.
\end{abstract}

\pacs{PACS Numbers: 75.10.Jm, 75.40.Mg }

]
\narrowtext
\section{Introduction}

The field of low dimensional quantum spin systems has been a focus in
condensed matter physics for about two decades since Haldane pointed out
the difference between the integer spin Heisenberg antiferromagnetic (HAF)
chains and the half-integer spin chains. By mapping the HAF chain to a non-linear
$\sigma$ model,\cite{haldane83}
Haldane conjectured that there is a finite gap between the ground state and the
low-lying excitation states for Heisenberg chains with integer spin,
while it is gapless for half-integer spin chains. This conjecture has
been verified by later analytic, numerical and experimental studies.

In recent years, the physics of spin ladders\cite{dagotto96} consisting of different 
number of coupled HAF spin-1/2 chains has attracted much attention. The spin ladders 
of even number coupled chains have a finite gap in low-energy spectrum and those of odd
number coupled chains are gapless.  The comparison of the quantum spin chains with the 
spin ladders suggests that there may be some relations between the two systems.  For 
the 2-leg ladders, White showed numerically that its ground
state can transform continuously to that of the S=1 spin chain no matter the
interchain coupling of the ladder is ferromagnetic or 
antiferromagnetic,\cite{white96} therefore they are in the same phase.  
While by considering a more general frustrated 2-leg ladder case, 
Wang identified two different phases, the Haldane phase and the
singlet phase, corresponding the ferromagnetic and the antiferromagnetic
interchain coupling respectively.\cite{wang00,zhu00} 

The effect of the magnetic field on the spin chains and the spin ladders
has been a topic of great interest. In the presence of a magnetic field, a spin-1/2
chain will not open a gap until the applied magnetic field is greater 
than the saturate field. While for a spin-1 chain, because of the existence
of the Haldane gap, there is a critical field where the gap is closed and
the system remains gapless from the critical field to the saturate field.
Below the critical field, there is a plateau at magnetization per site
$m$=0.  For those chains with higher spins, bond-alternating 
or other additional interactions, and the spin ladders, the magnetization process 
may also display plateaus at non-zero $m$.\cite{hida94,sakai98,cabra98,gier00}
It is shown\cite{oshikawa97} that for a spin chain with periodic structures,
in a uniform magnetic field the magnetization plateaus may exist
at the magnetization per site $m$ when the condition 
\begin{equation}
n(S-m) = \mbox{integer}
\label{srule}
\end{equation}
is satisfied. Here $n$ is the period of the ground state and $S$ is
the site spin. For ladder systems, the condition (\ref{srule}) needs some slight
modification. \cite{cabra98} The simplest case where a non-zero $m$ magnetization 
plateau may exist is the traditional translationally invariant S=3/2  spin chain, 
with n=1, and m=1/2 satisfying Eq. (\ref{srule}).  But detailed analyses have shown
that there is no plateau in this case \cite{oshikawa97} except when single-ion 
anisotropy $D$ is present and $D >$ 0.93.\cite{sakai98}  

Considerable attention has been given to mixed spin chains in recent years.
Materials with mixed ($S_1, S_2)$ chain structures have been synthesized. \cite{e1,e2}
Both types of excitations, gaped and gapless, in the low-energy spectrum of the
ferrimagnetic quantum alternating spin (1,1/2) chain have been obtained.
\cite{kolezhuk97,brehmer97,pati97,yamamoto98,wu99}
Magnetization plateau behavior in such systems \cite{yamamoto00} and in 
coupled alternating spin chains also has been studied. \cite{langari00}

In this paper, we present the results of the low-energy properties and the
magnetization plateau behavior of a mixed spin ladder consisting
of a spin-1/2 chain coupled with a spin-1 chain
\begin{equation}
H=\sum_{i=1}^N ( J_1 {\bf s}_i \cdot {\bf s}_{i+1}
                 +J_2 {\bf S}_i \cdot {\bf S}_{i+1} 
                 +J^\prime {\bf S}_i \cdot {\bf s}_i )
\label{ham}
\end{equation}
using the density matrix renormalization group (DMRG) method.\cite{dmrg}
Here ${\bf S}$ and ${\bf s}$ denote the spin operators on the spin-1 and 
spin-1/2 chains.  $N$ is the number of translational invariant cell ($S$,$s$)=(1,1/2). 
The number of total lattice sites is $L=2 N$.  We consider the case of antiferromagnetic 
intrachain coupling $J_1$, $J_2 > $  0 in this work.  To simplify the discussion, we 
fix $J_1 + J_2$=2 and make $J_1$ change its value from 0 to 2.  The interchain coupling
$J^\prime$ can be ferromagnetic or antiferromagnetic.  It is of considerable interest
to examine the effect of the nature of the interchain coupling on the properties of the 
mixed spin ladder, especially the possibility of introducing new phases as in the 
case of the spin-1/2 isotropic ladder. \cite{wang00,zhu00}

\begin{figure}[ht]
\epsfxsize=3.3 in\centerline{\epsffile{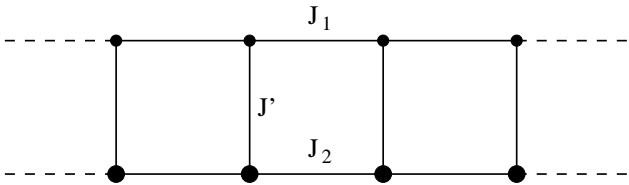}}
\vspace{0.2cm}
\caption[]{ The illustration of the 2-leg mixed spin ladder. Large (small) circles
denote S=1 (s=1/2) spin.  Intrachain coupling $J_1$, $J_2 >$  0.  }
\label{fig1}
\end{figure}

The mixed spin ladder is illustrated in Fig. \ref{fig1}.  From intuitive point of view,
when the interchain coupling is strongly ferromagnetic this mixed spin ladder is expected 
to be equivalent to the spin-3/2 Heisenberg chain, which is gapless in the low-energy 
spectrum with a spin wave velocity $\nu$ =3.87 and belongs to the same universality 
Haldane class as the S=1/2 Heisenberg chain. \cite{hallberg96} 
Our calculated low-energy properties provide strong support for this view.  
When the interchain coupling is antiferromagnetic, the calculated low-energy spectrum
does not show clear evidence for the equivalence.

The magnetic plateau study shows that the $J^\prime >$0 and $J^\prime <$0
phases behave differently.  In the presence of a uniform magnetic field $H$, 
an additional term
\begin{equation}
H^\prime = -h \sum_i (S_i^z+s_i^z)
\label{magham}
\end{equation}
needs to be added to Hamiltonian (\ref{ham}).  
When $J_1$ and $J^\prime$ equal zero, there is a m=1/2 magnetization
plateau from magnetic field H=0 to the critical field of the spin-1 chain,
which is related to the partially magnetized valence-bond-solid (VBS) state of the 
S=3/2 spin chain.\cite{oshikawa97,affleck87,affleck88,oshikawa92}
In our calculation, besides this magnetization plateau phase, we also have
identified another magnetization plateau phase when the interchain coupling is strongly
antiferromagnetic. We have calculated the magnetization distribution on the
two chains at $m=1/2$. In the first plateau phase, the magnetization is mainly
from the spin-1/2 chain as expected. But in the second plateau phase,
the magnetization comes mainly from the spin-1 chain, while the spins are antiparallel 
to the applied field in the spin-1/2 chain.

We will show the calculated low-energy properties of the 2-leg mixed spin ladder
in Sec. \ref{lowenergy}.  The magnetization plateau behavior and the phase diagram
are discussed in Sec. \ref{plateau}.  Finally, in Sec. \ref{conclusion}, we 
present the conclusion of the calculated results and some discussion on the
2-leg mixed spin ladders with general mixed spin cell ($S$, $s$).

\section{Low Energy Properties}
\label{lowenergy}

The ground state and low-energy excitation states of Hamiltonian (\ref{ham})
are calculated using DMRG by keeping 400 states. We use the open boundary
condition (OBC) and calculate up to the cell length N=60. The largest truncation
error is of the order of $10^{-7}$.

We  calculated the energies and correlation functions of the lowest state
in $S^z_{tot}$=0, 1, 2, 3 sectors for different $J_1$, $J_2$ and $J^\prime$.
For all the parameters we calculated, the low-energy spectrum is gapless 
at the thermodynamic limit and except for those very small $J^\prime$, it is
similar for different $J_1$, and $J_2$.  In the following we present the
calculated results of the excitation energy of low-lying states, the spin excitation
in the system, and the spin-spin correlation functions for the 2-leg mixed
spin ladder with $J_1=J_2=1$.

\begin{figure}[ht]
\epsfxsize=2.3 in\centerline{\epsffile{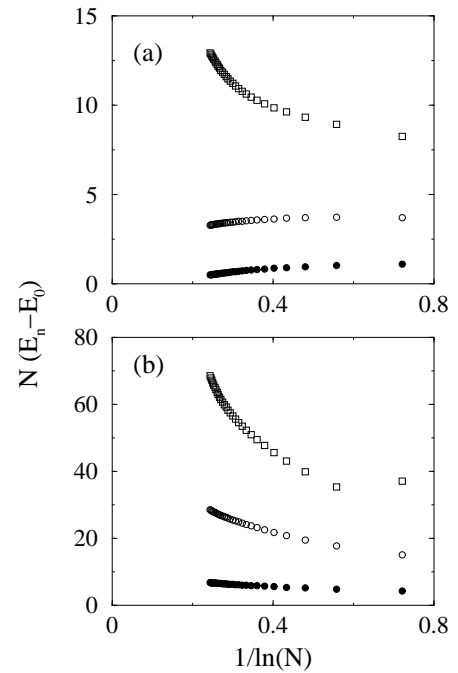}}
\vspace{0.2cm}
\caption[]{Excitation energy times the number of cells $N(E_n-E_0)$ for n=1 
(filled circle), n=2 (open circle) and n=3 (open square) are plotted versus
$1/\ln N$ for $J_1$=1.0 and (a) $J^\prime$=$-$4, (b) $J^\prime$=4.}
\label{lowen}
\end{figure}

The excitation energy of several lowest states times cell length $N (E_n-E_0)$ 
as a 
function of $1/\ln{N}$ for $J^\prime=-$4 and 4 are shown in Fig. \ref{lowen}(a) and (b)
respectively.  Here the energy of the lowest state for $S^z_{tot}=k$ sector is denoted as 
$E_k$.  For $J^\prime=-$4, the results are nearly identical to that of the spin-3/2 
chain. \cite{lou00,qin95}  Because we use OBC in the calculation, there is a s=1/2 free 
end spin which can be understood by valence bond solid picture.
\cite{affleck87,qin95,ng94,kim00} 
The $S^z_{tot}$=0 and $S^z_{tot}$=1 state will be degenerate 
in the thermodynamic limit. The energy spacing of the two quasi-degenerate ground states 
scales as $1/(N\ln N)$, exactly same as that for a pure spin-3/2 chain.\cite{lou00,qin95} 
For $J^\prime$=4, there is no end spin.  Here
the $S^z_{tot}$=0 state is the only ground state in the thermodynamic
limit, and the first excitation state at finite chain length is a true 
excitation state. The energy spacing between this state and the ground
state no longer scales as $1/(N\ln N)$.

\begin{figure}[ht]
\epsfxsize=3.3 in\centerline{\epsffile{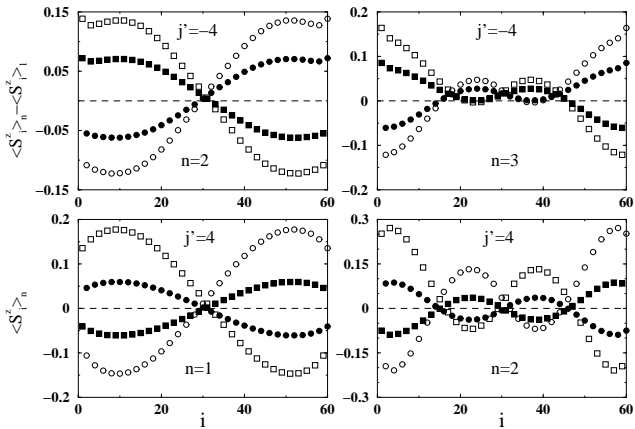}}
\vspace{0.2cm}
\caption[]{The excitation scheme $\langle S^z_i\rangle_n - \langle S^z_i \rangle_1$
(for $J^\prime=-$4) $\langle S^z_i \rangle_n$ (for $J^\prime$=4) on spin-1/2 chain 
(filled circle and square) and the spin-1 chain (open circle and square) for N=60.
The circles are for the even $i$ sites, while the squares are for the odd $i$ sites.}
\label{sz}
\end{figure}

The difference of the expectation value of the $z$ component of each spin site in the 
excitation states to the ground state is shown in Fig. \ref{sz}. For $J^\prime$=4, 
due to the reflection symmetry of the system, the expectation value $\langle S^z_i \rangle$ 
for $S^z_{tot}=0$ state is zero for all $i$, we show 
$\langle S^z_i \rangle$ directly. For $J^\prime=-$4, considering the existence of the 
end state, we show $\langle S^z_i \rangle_n - \langle S^z_i \rangle_1$.  These differences 
exhibit spin-wave-like feature of the excitations.  It is noticed that
the excitation of the spin-1 part and that of the spin-1/2 part are in phase for 
$J^\prime=-$4 but out of phase for $J^\prime$=4.  There is also clear difference between
the two cases near the chain end.

\begin{figure}[ht]
\epsfxsize=3.3 in\centerline{\epsffile{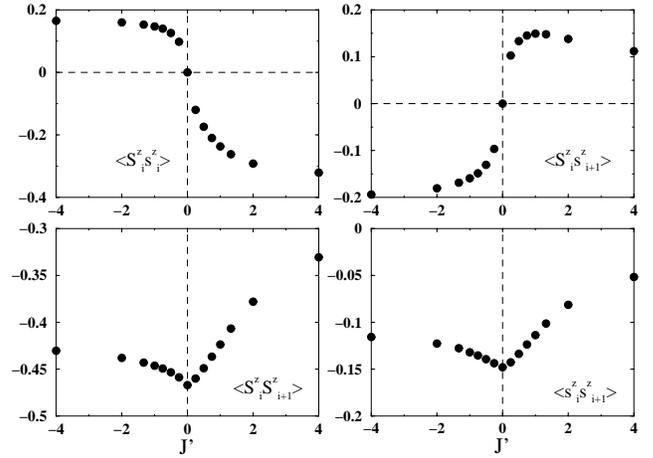}}
\vspace{0.2cm}
\caption[]{The dependence of the four neighboring correlation functions on the 
interchain coupling $J^\prime$ for $J_1 =1.0$.}
\label{corr}
\end{figure}

In Fig. \ref{corr}, we show the correlation between neighboring site spins as a function
of $J^\prime$.  The trends of these results are easy to understand by considering the
proper limiting cases.  When $J^\prime \to -\infty$, $\langle S^z_i s^z_i \rangle$ 
approaches 1/6, showing that the cell composed by the nearest-neighbor (1,$\frac{1}{2}$) 
spin pair behaves like a S=3/2 spin;
when $J^\prime \to \infty$, it approaches $-$1/3, the (1,$\frac{1}{2}$) spin pair is
coupled as a spin singlet.  The correlation $\langle S^z_i s^z_{i+1} \rangle$
decreases monotonically to a constant when $J^\prime \to -\infty$, whereas
when $J^\prime$ is positive, it first increases and reaches its maxim at 
$J^\prime \sim 1$ and then it decreases to zero for very large $J^\prime \to \infty$.
The two intrachain correlation functions  $\langle S^z_i S^z_{i+1} \rangle$
and $\langle s^z_i s^z_{i+1} \rangle$ reduce their magnitude when $J^\prime$ move away
from zero. For $J^\prime \to -\infty$, they will decrease to non-zero constants while 
for $J^\prime \to \infty$, they will decrease to zero. 

We believe there is a phase transition between the strongly 
ferromagnetic rung coupling region and the strongly 
antiferromagnetic rung coupling region.  The existence of the end 
free spin reveals the VBS topological order in the 
ferromagnetic rung coupling region.  Such a topological order
is absent in antiferromagnetic rung coupling region.  Therefore 
a quantum phase transition should occur.\cite{kim00}  We will 
further explore this phase transition in a future study.

\section{Magnetization Plateau}
\label{plateau}

For a gapless system, the conformal field theory\cite{cardy84,affleck86}
shows that the upper and lower critical magnetic field $H^+$ and $H^-$
where the ground state of the system under magnetic field has 
magnetization $m=M/N$ is\cite{sakai98},
\begin{eqnarray}
H^+_m &=& \lim_{N \to \infty} [E(N,M+1)-E(N,M)] \nonumber \\
H^-_m &=& \lim_{N \to \infty} [E(N,M)-E(N,M-1)].
\end{eqnarray}
Here $E(N,I)$ is the lowest energy in the subspace of an N-cell system with the
$z$ component of the total spin $\sum_i (S^z_i + s^z_i) = I$.
If $H^+_m = H^-_m = H_m$,  there is no plateau at magnetization $m$
in the thermodynamic limit and when the applied magnetic field
is $H_m$, the system has magnetization $m$.
If $H^+_m$ and $H^-_m$ are not equal, there will be a magnetization plateau
at $m$ with a width $D_m = H^+_m - H^-_m$.
By calculating the width of a length-$N$ system:
\begin{equation}
D_m(N) = E(N,M+1)+E(N,M-1) -2 E(N,M),
\end{equation}
$D_m$, the plateau width at magnetization $m$ can be
obtained by considering the infinite length limit
\begin{equation}
D_m = \lim_{N \to \infty} D_m(N).
\label{width}
\end{equation}

According to Eq. (\ref{srule}), Hamiltonian (\ref{ham}) may have a plateau
at magnetization per cell m=1/2. Using DMRG, we have calculated
$D_{\frac{1}{2}}(N)$ of this system and obtained two different plateau phases.

\subsection{$J^\prime$=0 limit}

Before presenting the DMRG results, we examine the limiting case of 
$J^\prime =0$.  In this case, the mixed spin ladder described by Hamiltonian (\ref{ham}) 
decouples into individual spin-1/2 and spin-1 chains. When a magnetic field is
applied to the decoupled system, the spin-1 chain will not be
magnetized until the field is larger than the critical field
$H^c_{1}=J_2 \Delta_{S=1}$, where $\Delta_{S=1}$ is the Haldane gap
of the pure spin-1 chain; while the gapless spin-1/2 chain will begin
magnetization as soon as the magnetic field is nonzero and
reach its saturate magnetization at $H^s_{\frac{1}{2}}=2 J_1$.
If $H^s_{\frac{1}{2}} < H^c_{1}$, when the spin-1/2 chain has been
saturated, the spin-1 chain has not been magnetized. Consequently, there 
will be a magnetization plateau between $H^s_{\frac{1}{2}}$
and $H^c_{1}$, with a magnetization per cell ($S$,$s$) of $m$=1/2.
This plateau will happen when $J_1$ is small and $J_2$ is large.  Since we fix 
$J_1+J_2 =2 $, we can obtain a critical point for $J_1$ where 
this plateau is diminished by taking 
\begin{equation}
J_2 \Delta_{S=1}=H^s_{\frac{1}{2}} = H^c_{1}=2 J_1.
\end{equation}
We have $J_1^c= \frac{2 \Delta_{S=1}}{2+\Delta_{S=1}}$.
The Haldane gap of the spin-1 chain $\Delta_{S=1}$  has been
calculated with high accuracy\cite{white93,golinelli94},
$\Delta_{S=1} \sim 0.41$, so $J_1^c \approx 0.34$.
When $J_1 < J_1^c$, there is a plateau at $m$=1/2 with
a width of $D_{1/2} = 2 \Delta_{S=1} - (2+\Delta_{S=1}) J_1$.

\subsection{DMRG results}

The energy levels $E(N,N/2)$, $E(N,N/2+1)$ and $E(N,N/2-1)$
of Hamiltonian (\ref{ham}) are calculated using DMRG for different
parameters $J_1$ and $J^\prime$.  Because of the open boundary
condition we use in calculation, for $J_1 < J_1^c$ and small
$J^\prime$ the edge states\cite{ng94,qin95} will also exist   
even at magnetization m=1/2.  As a result, the lowest state
in the $N/2+1$ sector may be the edge state. Considering this,
we target at least two states for the $E(N,N/2+1)$ sectors,
and the plateau behavior corresponds to the 
second state of this sector $E_2(N,N/2+1)$. 

We check the accuracy of our results by examining the case of $J_1=0$ and $J^\prime=0$, 
where the largest numerical error is expected for the DMRG calculation.
In this case, the decoupled spin-1/2 sites are free spins.  The m=1/2 plateau begins at 
zero and lasts to the Haldane gap of the spin-1 chain, $J_2 \Delta_{s=1} \sim $ 0.82. 
Our DMRG calculation gives $E_2(N,N/2+1)-E(N,N/2) \sim 0.832$. 

\begin{figure}[ht]
\epsfxsize=3.3 in\centerline{\epsffile{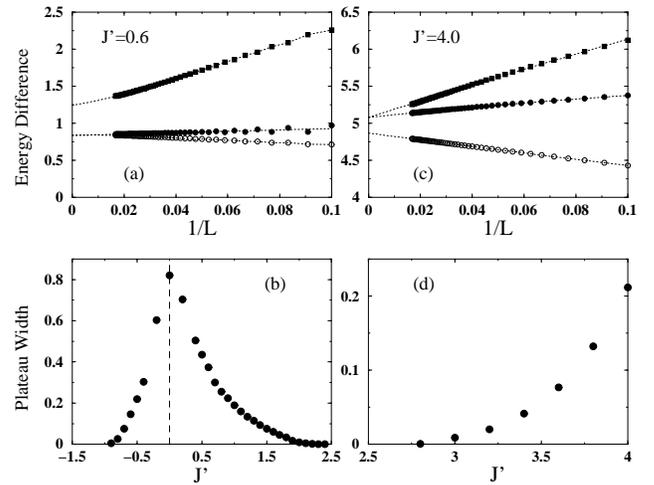}}
\vspace{0.2cm}
\caption[]{ Energy difference ((a) and (c))
$H^+_{\frac{1}{2} a}$ (filled circle),
$H^-_{\frac{1}{2}}$ (open circle),
$H^+_{\frac{1}{2} b}$ (filled square)
at m=1/2 and the plateau width ((b) and (d))
in the two plateau phases.
 }
\label{plat}
\end{figure}

When $J^\prime$ deviates from zero while $J_1$ still is zero, 
the width of the plateau will decrease. We show 
\begin{eqnarray}
H^+_{\frac{1}{2} a} & = & E(N,N/2+1)-E(N,N/2), \nonumber \\
H^+_{\frac{1}{2} b} & = & E_2(N,N/2+1)-E(N,N/2), \nonumber \\
H^-_{\frac{1}{2}}   & = & E(N,N/2)-E(N,N/2-1) 
\end{eqnarray}
at $J_1=0$ and $J^\prime=0.6$ in Fig.\ref{plat} (a).
Clearly $H^+_{\frac{1}{2} a}$ and $H^-_{\frac{1}{2}} $
converge to the same value in the thermodynamic limit. 
The plateau width is the difference of $H^+_{\frac{1}{2} b}$
and $H^-_{\frac{1}{2}} $, which is about 0.42, smaller
than that of the $J^\prime$=0 case.  For $J^\prime < $0,
the width decreases faster than in the $J^\prime > $0 case. At critical points
$J^\prime_{c-}(J_1=0) \sim 0.85$ and $J^\prime_{c+}(J_1=0) \sim 2.2$, the width
decreases to zero and the plateau vanishes.  
In these cases, the width of the plateau can also be
obtained from the difference of the $E(N,N/2+1)$
and $E_2(N,N/2+1)$, $D_m = \lim_{N \to \infty} [E_2(N,N/2+1)-E(N,N/2+1)]$.
This definition gives $D_m$ with higher accuracy.  The dependence of the width 
of the plateau on $J^\prime$ for $J_1=0$ is shown in Fig.\ref{plat} (b).

For larger $J^\prime > J^\prime_C$, we find another plateau phase. 
The values of $H^+_{\frac{1}{2} a}$ , $H^+_{\frac{1}{2} b}$ ,
and $H^-_{\frac{1}{2}} $ at $J_1=0$ and $J^\prime=4.0$ are shown in Fig.\ref{plat} (c).
For this larger $J^\prime$, $E_2(N,N/2+1)$ is not
an edge state. The non-zero magnetization plateau is
obvious. The width of the magnetization plateau for $J_1$=0
and $J^\prime > J^\prime_C(J_1=0) \sim 2.8$
is shown in Fig. \ref{plat} (d).  

\begin{figure}[ht]
\epsfxsize=2.3 in\centerline{\epsffile{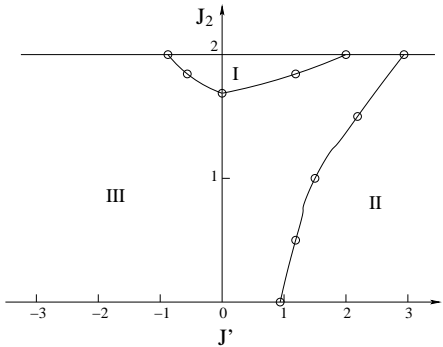}}
\vspace{0.2cm}
\caption[]{ The magnetization plateau phase diagram, the open 
circles are the critical points obtained by DMRG calculations.}
\label{phase}
\end{figure}

\begin{figure}[ht]
\epsfxsize=2.8 in\centerline{\epsffile{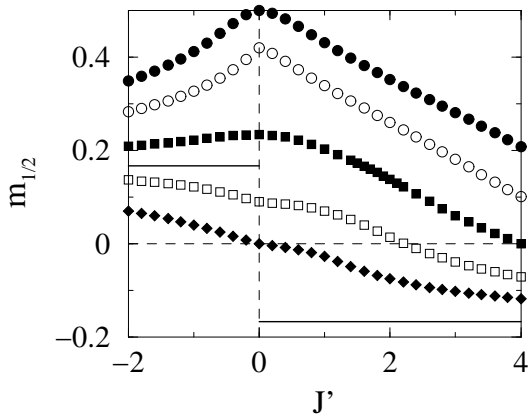}}
\vspace{0.2cm}
\caption[]{Spin-1/2 chain magnetization $m_{\frac{1}{2}}$ at $m$=1/2 
for $J_1$ = 0 (filled circle), 0.5 (open circle), 1.0 (filled square), 1.5 
(open square), and 2.0 (filled diamond).  The solid lines are the spin-1/2 chain 
magnetization at $J^\prime \rightarrow \pm \infty$ limit.}
\label{submag}
\end{figure}

We have carried out systematic DMRG calculations to determine the magnetization plateau
phase diagram shown in Fig. \ref{phase}.  An examination of the calculated results
indicate that there are two types of plateau phases corresponding to different
magnetic structures in the ladder.  In phase I, the magnetization is due to 
the partially polarized s=1/2 spins. In phase II, the spin-1/2
sites and spin-1 sites are strongly coupled.  In phase III, where no plateau is
found, the system is in a phase similar to that of a pure spin-3/2 chain. 
The magnetic features of the two plateau phases in Fig. \ref{phase} can be
clearly seen in Fig. \ref{submag}, which shows the magnetization distribution 
on the spin-1/2 chain defined as
\begin{equation}
m_{\frac{1}{2}} = \frac{1}{N} \sum_i \langle s^z_i \rangle
\end{equation}
The magnetization distribution on the spin-1 chain can be derived from 
$m_1 = m - m_\frac{1}{2}$, where $m$ is the total magnetization of the ladder.

To understand the trends shown in Fig. \ref{submag}, it is instructive to examine the
two limiting cases.  When $J^\prime \to -\infty$, $m_\frac{1}{2} \to 1/6$ and 
$m_1= m-m_\frac{1}{2} \to 1/3$, and the magnetizations on the two
chains are half of their saturate values. In this limit, each cell ($1$,$1/2$) is in 
$|3/2,1/2 \rangle = \frac{1}{\sqrt{3}} (\sqrt{2} |1,1;1/2,-1/2 \rangle
+ |1,0;1/2,1/2 \rangle )$ state when $m$=1/2, and the 1/2-spin
and the corresponding 1-spin bond together to form a 3/2-spin.
When $J^\prime \to \infty$, $m_\frac{1}{2} \to -1/6$, that
is, the orientation of the spin-1/2 partition will be antiparallel
to the applied magnetic field.  Here each ($1$,$1/2$) cell is in 
$|1/2,1/2 \rangle = \frac{1}{\sqrt{3}} (|1,1;1/2,-1/2 \rangle
- \sqrt{2} |1,0;1/2,1/2 \rangle )$ state when  $m$=1/2.

\begin{figure}[ht]
\epsfxsize=2.8 in\centerline{\epsffile{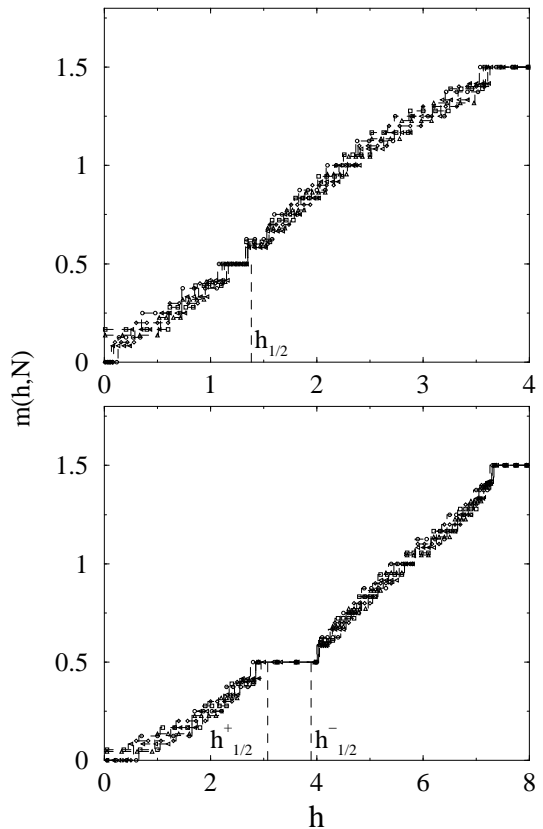}}
\vspace{0.2cm}
\caption[]{Magnetization process of short chains with N=8, 9, 10, 11, and 12
for $J_1$=1.0 and (a) $J^\prime=-$1.0, (b) $J^\prime$=3.0. 
The magnetization plateau at m=1/2 is obvious for $J^\prime$=3.0.}
\label{magpro}
\end{figure}

We also have studied the magnetization process by examining short chains.
We show the typical plateau and no-plateau process in 
Fig.\ref{magpro}(a) and Fig.\ref{magpro}(b) with $J_1$=1.0, 
$J^\prime = -1.0$ and $3.0$ respectively.
For $J^\prime=3.0$ (phase II in Fig. \ref{phase}), the existence of the magnetization 
plateau at $m=\frac{1}{2}$ is clear. When the chain length grows,
$h^+_{1/2}$ becomes larger and $h^-_{1/2}$ becomes smaller, 
but they converge to different values for the infinite chain.
For $J^\prime=-$1.0 (phase III in Fig. \ref{phase}),  both $h^+_{1/2}$ and $h^-_{1/2}$ 
increase with chain length, but $h^+_{1/2}$ increase faster.
They eventually converge to the same value $h_{1/2}$, yielding no plateau in this case. 

For $J_1=J_2=1$ with magnetization 1/2, it is known by perturbation that 
there is no spin gap for strong ferromagnetic rung coupling region, 
and there is a finite spin gap for strong antiferromagnetic rung 
coupling region.  We have reported calculations for the intermediate rung 
coupling region by DMRG method in this section.  We find that there is only one 
quantum phase transition at which the spin gap appears when we 
increase the rung coupling $J'$ from a negative magnitude to a
big and positive one.  Beside this quantum phase transition between
plateau phase II and plateauless phase III, the phase transition
between plateau phase I and plateauless phase III is also a 
continuum phase transition due to spin gap closing.  A more precise
understanding of the gap generation and magnetization plateaus 
in the phase diagram found in this section can be given by
bosonization discussions.  We expect that further bosonization studies
will provide an interpretation of the plateau phase II beyond the 
obvious origin of its plateau.

\section{Conclusion and Discussion}
\label{conclusion}

We have calculated the low-energy properties and magnetization plateau
behavior of a 2-leg mixed spin ($S$,$s$)=(1,1/2) ladder.  From the low-energy 
properties, we conclude that for strong ferromagnetic interchain coupling, the 
2-leg mixed spin ladder is in the same phase as the 
spin-3/2 chain. There is a magnetization plateau for large enough
positive $J^\prime$ but no plateau is found for large negative
$J^\prime$. This suggests that at large $J^\prime > $  0, the
ladder would behave differently from the spin-3/2 chain.

We have obtained the magnetization plateau phase diagram of Hamiltonian
(\ref{ham}) and identified two types of plateau phases. In the
partially polarized spin-1/2 phase, the total magnetization comes mainly from
the spin-1/2 chain, while the spin-1 chain is only polarized very weakly. 
In another plateau phase, because of the antiferromagnetic interchain
coupling, the orientation of the spin-1/2 chain is antiparallel to the magnetic field, 
and the spin-1/2 chain and the spin-1 chain are coupled together.

Finally, we discuss some qualitative trends for 
general mixed spin ($S$,$s$) ladders.  The basic structure of the phase diagram shown 
in Fig. \ref{phase} is expected to remain valid.  For higher spins ($S$,$s$), the 
non-plateau region phase III for (1,1/2) may also become a plateau phase.
If $S$ and $s$ are both half-integers, there will be no partially polarized plateau phase.
If $S$ and $s$ are one integer and one half-integer, then the low
energy spectrum is gapless and there is no plateau at m=0. In other
cases, the low-energy spectrum is gaped, with a m=0 plateau.

\acknowledgments
This work was support in part by the Department of Energy at the University of Nevada,
Las Vegas. S. Qin was partially supported by the Chinese Natural Science Foundation.
J. Lou would like to thank Prof. Xiaoqun Wang for useful discussions.


\begin{references}
\bibitem{haldane83} F. D. M. Haldane, Phys. Rev. Lett. {\bf 50}, 1153 (1983);
                    F. D. M. Haldane, Phys. Lett. {\bf 93A}, 464 (1983).

\bibitem{dagotto96} E. Dagotto and T.M. Rice, Science {\bf 271}, 618 (1996).

\bibitem{white96}  S. R. White, Phys. Rev. B {\bf 53}, 52 (1996).

\bibitem{wang00} Xiaoqun Wang,  Mod. Phys. Lett. B {\bf 14}, 327 (2000);
                 also in {\it Density-Matrix Renormalization}, 
                 edited by I. Peshel, X. Wang,
                 M. Kaulke and K. Hallberg, Lecture Notes in Physics 
                 (Springer, New York, 1999).

\bibitem{zhu00}  N. Zhu, X. Wang and C. Chen,
                 Phys. Rev. B {\bf 63}, 012401 (2001).

\bibitem{hida94} K. Hida, J. Phys. Soc. Jpn. {\bf 63}, 2359 (1994).

\bibitem{sakai98} T. Sakai and M. Takahashi,
                  Phys. Rev. B {\bf 57}, R3201 (1998).

\bibitem{cabra98} D. C. Cabra, A. Honecker and P. Pujol,
                  Phys. Rev. B {\bf 58}, 6241 (1998).

\bibitem{gier00} J. de Gier and M. T. Batchelor,
                 Phys. Rev. B {\bf 62}, R3584 (2000).

\bibitem{oshikawa97} M. Oshikawa, M. Yamanaka adn I. Affleck,
                 Phys. Rev. Lett. {\bf 78}, 1984 (1997).

\bibitem{e1} M. Hagiwara, K. Minami, and K. Kindo, J. Phys. Soc. Japan {\bf 67}
, 2209 (1998).

\bibitem{e2} M. Verdaguer, A. Gleizes, J. P. Renard, and J. Seiden, Phys. Rev. 
B {\bf 29},
5144 (1984).

\bibitem{kolezhuk97} A. K. Kolezhuk, H.-J. Mikeska, and S. Yamamoto,
                    Phys. Rev. B {\bf 55}, R3336 (1997).

\bibitem{brehmer97} S. Brehmer, H.-J. Mikeska and S. Yamamoto,
                    J. Phys. Cond. Matt. {\bf 9}, 3921 (1997).

\bibitem{pati97}  S. K. Pati, S. Ramasesha, and D. Sen,
                  Phys. Rev. B {\bf 55}, 8894 (1997).

\bibitem{yamamoto98} S. Yamamoto, T. Fukui, K. Maisinger, and U. Sch\"ollwock,
                     J. Phys. Cond. Matt. {\bf 10}, 11033 (1998).

\bibitem{wu99} C. Wu, B. Chen, X. Dai, Y. Yu, and Z.-B. Su,
               Phys. Rev. B {\bf 60}, 1057 (1999).

\bibitem{yamamoto00} S. Yamamoto and T. Sakai,
               Phys. Rev. B {\bf 62}, 3795 (2000).

\bibitem{langari00} A. Langari and M. A. Martin-Delgado,
               Phys. Rev. B {\bf 62}, 11725 (2000).

\bibitem{dmrg} S. R. White, Phys. Rev. Lett. {\bf 68}, 3487 (1992);
                Phys. Rev. B {\bf 48}, 10345 (1993).

\bibitem{hallberg96} K. Hallberg, X. Q. G. Wang, P. Horsch, and A. Moreo,
                    Phys. Rev. Lett. {\bf 76}, 4955 (1996).

\bibitem{affleck87} I. Affleck, T. Kennedy, E. H. Lieb, and H. Tasaki,
                Phys. Rev. Lett. {\bf 59}, 799 (1987).

\bibitem{affleck88} I. Affleck, T. Kennedy, E. H. Lieb, and H. Tasaki,
                    Commun. Math. Phys. {\bf 115}, 477 (1988).

\bibitem{oshikawa92} M. Oshikawa, 
                     J. Phys. Cond. Matt. {\bf 4}, 7469 (1992).

\bibitem{lou00} J. Lou, J. Dai, S. Qin, Z. Su, and L. Yu,
                Phys. Rev. B {\bf 62}, 8600 (2000).

\bibitem{qin95} S. Qin, T.-K. Ng, and Z.-B. Su,
                Phys. Rev. B {\bf 52}, 12844 (1995).

\bibitem{ng94} T.-K. Ng, Phys. Rev. B {\bf 50}, 555 (1994).

\bibitem{kim00} Eugene H. Kim, G. Fath, J. Solyom and D. J. Scalapino,
                Phys. Rev. B {\bf 62}, 14965 (2000).

\bibitem{cardy84} J. L. Cardy, J. Phys. A {\bf 17}, L385 (1984);
                 H. W. Bl\"ote, J. L. Cardy, and M. P. Nightingale,
                 Phys. Rev. Lett. {\bf 56}, 742 (1986).
               
\bibitem{affleck86} I. Affleck, Phys. Rev. Lett. {\bf 56}, 746 (1986).

\bibitem{white93} S. R. White and D. A. Huse, Phys. Rev. B 48, 3844 (1993).

\bibitem{golinelli94} O. Golinelli, Th. Jolicoeur, and R. Lacaze, 
                 Phys. Rev. B 50, 3037 (1994).
\end{references}
\end{document}